\documentclass[prl,%12pt,
twocolumn,
showpacs,showkeys,floatfix]{revtex4}\usepackage{graphicx}
\usepackage{bm}
\usepackage{latexsym}
\newcommand{\beq}{\begin{equation}}
\newcommand{\eeq}{\end{equation}}
\newcommand{\bc}{\begin{center}}
\newcommand{\ec}{\end{center}}
\newcommand{\eeqa}{\end{eqnarray}}
\newcommand{\beqa}{\begin{eqnarray}}
\newcommand{\no}{\noindent}
\newcommand{\pa}{\partial}

\newcommand{\ra}{\rightarrow}

\newcommand{\al}{\alpha}
\newcommand{\be}{\beta}

\newcommand{\de}{\delta}

\newcommand{\ep}{\epsilon}

\newcommand{\la}{\lambda}

\newcommand{\si}{\sigma}

\newcommand{\ph}{\phi}

\newcommand{\om}{\omega}
\newcommand{\ed}{\end{document} }

\begin{document}

\raggedright
\parindent3em

\title{Lorentz covariant and gauge invariant description of orbital and spin angular momentum and the non-symmetric energy momentum tensor}
\author{Richard T. Hammond}
\email{rhammond@email.unc.edu}
\affiliation{Department of Physics,
University of North Carolina at Chapel Hill,
and the
Army Research Office
Research Triangle Park, North Carolina, 27703}

\date{\today}


\begin{abstract}
There is a thriving controversy about the correct mathematical description of spin angular momentum and orbital angular momentum. The reason this is such an unsettled question is that the results are usually not gauge invariant, and therefore not physical. Starting from covariant expressions, a gauge invariant separation of orbital and spin angular momentum for electrodynamics is presented. This results from the non-symmetric canonical energy momentum tensor of the electromagnetic field. The origin of the difficulty is discussed and a covariant, gauge invariant spin vector and orbital angular momentum vector are derived. The longstanding paradox concerning the spin angular momentum of a plane wave finds a natural solution.
\end{abstract}
\maketitle

Recently the interaction of spin and orbital angular momentum has been observed,\cite{dash} spin dependent forces were measured using a cantilever apparatus,\cite{ant} and electrons in a circular orbit were observed to create beams with orbital angular momentum in addition to spin.

Much of the recent work was inspired by the paper by Allen {\em et al} in 1992, yet controversies about the
orbital versus spin angular momentum of light continue to thrive. To see  into the heart of the controversy, we begin with some textbook physics. The momentum density of the electromagnetic field in vacuum is

\beq
{\bf P} = \frac{1}{4\pi c}{\bf E}\times{\bf B}
\eeq
so the total angular momentum of the electromagnetic field with respect to  the ${\bf r}=0$ axis is the volume integral,

\beq\label{j}
{\bf J}=\int  {\bf r}\times{\bf P}d^3x
.\eeq
Using vector identities and eliminating surface terms this may be written
\
\beq\label{jsplit}
{\bf J}= \frac{1}{4\pi c}\sum_n\int  
\left(
E_n({\bf r}\times {\bm \nabla})A_n-
{\bf E}\times{\bf A}
\right)d^3x
.\eeq
The first term, which depends on $r$, is called the orbital angular momentum, and the second part, independent of the coordinates, a term intrinsic to the field, is called the spin.

However, the spin term is not gauge invariant and neither is the orbital angular momentum, so how can these be physical?  The approach sometimes taken   is simply to use the transverse fields by breaking up the electric field into its rotational and irrotational parts,\cite{enk} but there is no physical justification for this and, in fact, only the total fields are physical and propagate at the speed of light.
Barnett {\it et. al.} have recently investigated this issue, and concluded, among other things, referring to the orbital and spin angular momentum, ``...neither alone is a true angular momentum.''\cite{barnett}

Another age old  issue has been the observation that, for a plane wave traveling in, say, the $z$ direction, since ${\bf E}\times {\bf B}$ is in the $z$ direction,  therefore from (\ref{j}), the $z$ component of the angular momentum is zero, which we know is not true from experiment.\cite{barnett} A plane circularly polarized wave does indeed have spin angular momentum. The extensive list of articles on these subjects is not presented here, but may be found in the above references, and elsewhere.\cite{leader}

There are two fundamental problems with the above. One is, the results are not gauge invariant, and therefore not physical, and the other is they are not Lorentz invariant, since a special frame was chosen. In the end we will choose the same frame, but it is better to start with a generally covariant approach, so let us consider the relativistic and covariant generalization of (\ref{j}).\cite{jac} This will also allow us to use  either the metric or canonical energy momentum discussed below.

The 4D generalization of ${\bf r}\times{\bf P}$ is defined as

\beq\label{jdef}
M^{\mu\nu\si}\equiv 
x^\mu T^{\nu\si}-x^\nu T^{\mu\si}
\eeq

\no where ${\bf r} =\{x^n\}$ (with $n=1,2,3$) and $x^0=ct$ and $T^{\mu\nu}$ is the energy momentum tensor, later taken to be that of the electromagnetic field. 
One might think conservation of angular momentum implies $\frac{d}{dt} M^{\mu\nu\si}=0$, but this is not a covariant statement, and we must be more careful.
The angular momentum tensor is defined as\cite{mtw}

\beq\label{jmunu}
J^{\mu\nu}=\int M^{\mu\nu\si}d\Sigma_\si
\eeq
where $d\Sigma_\si$ are the ``surface" elements in four dimensions.\cite{lan}
For example, over a constant time hypersurface this becomes

\beq
J^{\mu\nu}=\int M^{\mu\nu0}dxdydz
.\eeq

Assuming $J^{\mu\nu}$ is conserved for a closed system, then
it is independent of which hypersurface (3D volume)  it is evaluated, so,

\beq\label{jj0}
J^{\mu\nu}_{\pa_a}-J^{\mu\nu}_{\pa_b}=0
\eeq
where ${\pa_a}$ and ${\pa_b}$ simply designate the hypersurfaces bounding the 4-volume, and therefore,

\beq
J^{\mu\nu}_{\pa_a}-J^{\mu\nu}_{\pa_b}=\oint M^{\mu\nu\si}d\Sigma_\si
\eeq
where the integral is over a closed 3-surface, so we may use Gauss's theorem in four dimensions,

\beq\label{j4}
J^{\mu\nu}_{\pa_a}-J^{\mu\nu}_{\pa_b}=\int d^4x M^{\mu\nu\si},_\si
.\eeq
Finally, from (\ref{jj0}) we have

\beq\label{dmz}
\int M^{\mu\nu\si},_\si d\Sigma_\si
=0
,\eeq
and so, since this holds for arbitrary volumes, $M^{\mu\nu\si},_\si =0$.
This is the well-known condition for conservation of angular momentum. These results will be used after deriving the energy momentum tensor.

Now consider the energy momentum tensor. There are two ways to derive the energy momentum tensor. One method, as found in many standard books\cite{jac}, yields the canonical energy momentum tensor derived via translational symmetry of the Lagrangian density, as shown by Noether. The main issue is that this derivation yields a non-symmetric tensor ($T^{\mu\nu} \neq T^{\nu\mu}$).  However, the energy momentum tensor may also be derived by considering variations of the metric tensor. With a symmetric metric tensor this gives a symmetric energy momentum tensor,

\beq\label{semt}
T^{\mu\nu}_m=\frac{1}{16\pi c}\left(g^{\mu \nu}F_{\al\be}F^{\al\be}-4F^{\mu\si}F^\nu_{ \ \si}\right)
\eeq
where $g^{\mu \nu}$ is the metric tensor, from now on taken to be that of Minkowski spacetime, $\eta^{\mu\nu}$, and
the electromagnetic field tensor is
\beq\label{f}
F_{\mu\nu}=A_{\nu,\mu}-A_{\mu,\nu}
.\eeq
However, with a non-symmetric metric tensor the energy momentum tensor is not symmetric.\cite{pap}\cite{ham}

Now let us find the relativistic generalization of (\ref{jsplit}).
Using (\ref{semt}), and assuming we are in a source free region so that $
F^{\mu\nu},_\nu=0$, so that $T^{\mu\nu},_\nu=0$ and also using $\pa x^\mu/ \pa x^\nu=\de^\mu_\nu$,
this may be written as

\begin{widetext}
\beq\label{xxx}
 16\pi  c M^{\mu\nu\si}=x^\mu\left(
 \eta^{\nu\si}F_{\al\be}F^{\al\be} -4F^{\nu\ph}A_\ph^{\ ,\si}
 \right)
 +(4x^\mu F^{\nu\ph}A^\si),_\ph
 -4F^{\nu\mu}A^\si -(\mu\leftrightarrow\nu)
 \eeq
 \end{widetext}
 where $(\mu\leftrightarrow\nu)$ means the same term as before it with $\mu$ and $\nu$ interchanged. Since this is really under an integral (see (\ref{jmunu}))
we see the second term on the right side is a surface term: Integrating over the four volume, and using Gauss' theorem, and assuming the surface terms go to zero on the boundary, this becomes,

\begin{widetext}
\beq\label{jsplitcov}
 16\pi c M^{\mu\nu\si}=x^\mu\left(
 \eta^{\nu\si}F_{\al\be}F^{\al\be} -4F^{\nu\ph}A_\ph^{\ ,\si}
 \right)
 -4F^{\nu\mu}A^\si -(\mu\leftrightarrow\nu)
. \eeq
\end{widetext}

This is the covariant generalization of (\ref{jsplit}), and the first term on the right side of (\ref{jsplitcov}) is called the orbital angular momentum and the second term is called the spin. As before the identification of orbital angular with the first term,  spin angular momentum with the second makes no sense since these terms are not gauge invariant. What's worse, the result (\ref{jsplitcov}), {\em as a whole}, is not gauge invariant. How is this, since we began with a gauge invariant expression?
Equally strange is,  letting  $x^\mu=0$ (say this is at $t=0$), from (\ref{jdef}) we see $M^{\mu\nu\si}=0$, but (\ref{jsplitcov}) shows we have a non-zero result. 

It is easy to trace along the derivation to see where things went haywire, it was after throwing off the surface term. That broke gauge invariance, and we see the reason for the contradiction about $x^\mu=0$ is, in obtaining (\ref{jsplitcov}) we used $x^\mu,_\nu=
\de^\mu_\nu$, but this is evidently false if $x^\mu=0$.
The surface term in (\ref{xxx}),
$(4x^\mu F^{\nu\ph}A^\si),_\ph$, was assumed to vanish on the surface.
Now, $F^{\nu\ph}$ is either a component of $\bf E$ or $\bf B$, say $\bf E$.
Also, if $E \sim A/r$ and the volume element goes like $r^2$, so the surface term $\sim r^2E^2$. Unless $E$ goes to zero faster than $1/r$, the surface term may not be thrown away. For a plane wave, $E \sim A$, it's even worse, and  this term cannot be set to zero.  
For a Laguerre-Gaussian beam, the radial dependence falls exponentially. Therefore, if the closed surface is a right cylinder with top, bottom, and face, then the contribution goes to zero on the face, but not on the top and bottom.
A better route to understanding angular momentum and spin is the application of the non-symmetric metric tensor.

First, let us review a textbook argument for assuming the energy momentum tensor to be symmetric. In the following we adopt the convention that the equations are under the integral. At the end, it is assumed the volume is arbitrary and at that point the integrand may be set to zero. The reason for this is that there will be surface terms we might throw away (although, as seen above, we must take care). So for example, from (\ref{dmz}) we have

\beq\label{jder}
M^{\mu\nu\si},_\si=
\left(x^\mu T^{\nu\si}-x^\nu T^{\mu\si}\right),_\si
,\eeq
which gives
\beq
M^{\mu\nu\si},_\si=
2T^{[\nu\mu]}-x^\nu T^{\mu\si},_\si+x^\mu T^{\nu\si},_\si=0
,\eeq
where $T^{[\mu\nu]}\equiv1/2(T^{\mu\nu}-T^{\nu\mu})$ is the antisymmetric part of the energy momentum tensor. Following (\ref{dmz}),  $M^{\mu\nu\si},_\si$ is set to zero, so that, if $T^{\mu\si},_\si=0$ (source free region),
then the antisymmetric part is zero and therefore the energy momentum tensor is symmetric.

The gaping flaw in this argument was pointed out long ago by Papapetrou.\cite{pap} Since (\ref{jdef}) contains $x^\mu$, from the outset it is the definition of orbital angular momentum. It does not contain spin terms (which are independent of $x^\mu$), so Papapetrou assumed

\beq\label{pjdef}
M^{\mu\nu\si}\equiv 
x^\mu T^{\nu\si}-x^\nu T^{\mu\si}+\la^{\mu\nu\si}
\eeq
so $M^{\mu\nu\si},_\si=0$,  interpreted as the conservation of {\em total }angular momentum, yields

\beq\label{lam}
\la^{\mu\nu\si}_{\ \ \ \ ,\si}= 2T^{[\mu\nu]}
 .\eeq
 In words, this us telling is the antisymmetric part of the energy momentum tensor is
 related to spin. This has also been recognized elsewhere \cite{bli} This is a fundamental result and is not restricted to electromagnetism.
 For example, in gravitation with a non-symmetric metric tensor
 we have\cite{ham}
 
\beq\label{tor}
S^{\mu\nu\si}_{\ \ \ \ ,\si}= KT^{[\mu\nu]}
 \eeq
where $S^{\mu\nu\si}$ is the torsion tensor and $K$ is the coupling constant.\cite{ham2}

 In the canonical electrodynamic approach, the energy momentum tensor may be written as

\beq\label{nsemt}
T^{\mu\nu}_{\mbox\small can}=T^{\mu\nu}_m-
\frac{1}{4\pi}F^{\mu\si}A^\nu ,_\si
.\eeq
With (\ref{lam}) we find

\beq\label{lam2}
\la^{\mu\nu\si}=\frac{1}{4\pi}\left(
F^{\mu\si}A^\nu-F^{\nu\si}A^\mu\right)
 .\eeq

It should be noted that in using the canonical energy momentum tensor in (\ref{pjdef}), we used $T^{\mu\nu}_{\mbox{\tiny{can}}},_\nu=0$, which is true provided
$(F^{\nu\si}A^\si,_\ph),_\si\ra0$, which is true if
$E^2\ra 0$ on the boundary, which is a much more benign condition than we had before ($r^2E^2\ra 0$). Still, it appears problematic for a plane wave, but it turns out this term is zero for a plane wave, as can be shown below. One should also note the canonical energy momentum tensor is conventionally symmetrized by assuming a term like $F^{\mu\si}A^\nu$ is thrown away on the boundary. But as seen before this is not always justified and, in general, {\em we must use the antisymmetric energy momentum tensor}.

Now we would like to define the angular momentum {\em vector}. For massive objects it is defined as

\beq
J_\xi=\frac{v^\ph}{2c}\ep_{\mu\nu\ph\xi}J^{\mu\nu}
\eeq
where $\ep_{\mu\nu\ph\xi}$ is the totally antisymmetric tensor, and $v^\ph$ is the velocity of the matter, but this makes no sense for light.\cite{mtw} Taking quantum mechanics as a guide, we know the velocity is  proportional to the momentum, and momentum is proportional to $k^\mu$, the wave vector, so let us consider, for monochromatic light (non-monochromatic light has been considered\cite{man}),

\beq\label{spindef}
J_\xi=\frac{c}{2\om}k^\mu\ep_{\al\be\mu\xi}J^{\al\be}
.\eeq
This can be broken into orbital and spin parts as $J_\xi= L_\xi +S_\xi$ respectively where, as always, the spin part does not contain the coordinate, so

\beq
S_\xi=\frac{c}{4\pi\om}k^\mu\ep_{\al\be\xi\mu}\int d\Sigma_\si F^{\al \si}A^\be
\eeq
and

\beq\label{lxi}
L_\xi=\frac{c}{4\pi\om}k^\mu\ep_{\al\be\mu\xi}\int d\Sigma_\si x^\al T_m^{\be\si}
\eeq

Under the gauge transformation $A_\al\ra A_\al+\la,_\al$, the $\la$ term can be converted to a surface integral which vanishes if $\la$ (and/or the fields) vanish on the hypersurface. This is much different than the conditions used above that led us so far astray). Thus we have a gauge invariant description of orbital (since the orbital angular momentum is already gauge invariant) and spin angular momentum.

Let us look at the spin vector in a hyperplane of constant time to show this definition makes sense.

\beq\label{sxi}
S_\xi=\frac{c}{4\pi\om}k^\mu\ep_{\al\be\xi\mu}\int dVF^{\be 0}A^\al
\eeq
where $dV = dxdydz$.
Consider a polarized wave, the potential of which is

\beqa\label{a}
A_\mu\equiv\{A_0, A_1, A_2, A_3\}\\ \nonumber
=\frac{ cE}{\om}\{0, \sin(kz-\om t), r \cos(kz-\om t),0\}
\eeqa
where $0  \leq r \leq 1$ represents the degree of circular polarization with $r=0$ being a plane polarized wave and $r=1$ a circularly polarized wave. With (\ref{f}) we find
$E_x=-E \cos\Upsilon$, $E_y=rE \sin\Upsilon$, $B_x=-r E \sin\Upsilon$, and $B_y=-E \cos\Upsilon$ where $\Upsilon = kz-\om t$.
In using (\ref{a}) the Lorentz gauge has been chosen, which we may do since we have gauge invariant results.
This represents a monochromatic wave propagating in the $z$ direction with  $c=\om/k$. Using (\ref{f}) and (\ref{a}) in (\ref{sxi}) we find

\beq\label{sz}
\frac {S_z}{V}=\frac{r}{4\pi \om}E^2
\eeq
where $V$ is the volume. If we assume $r=1$ and the wave consists of $n$ photons per unit volume, then the total spin density should be $n\hbar$ since each photon contributes $\hbar$. The energy density of the wave, $u$, is 
$u=(E^2+B^2)/8\pi =E^2/4\pi$, and with (\ref{sz})

\beq
u=n\hbar\om
\eeq
which is the correct result giving the energy density in terms of the photon number density. This result also resolves the age old issue described above in a natural and simple way. Using (\ref{a}), we may also show $T^{\mu\nu}_{\mbox{\tiny{can}}},_\nu=0$ is zero, as claimed above (although the surface term examined previously, $(4x^\mu F^{\nu\ph}A^\si),_\ph$, is not zero for a plane wave).

However, the orbital part of the angular momentum vector, $L_\xi$ is different than (\ref{jsplit}), so we should consider ways to test these results. For example, once again on a hypersurface of constant time we find (\ref{lxi}) reduces to

\beq\label{lf}
{\bf L}=\int dV {\bf r}\times{\bf P}
\eeq
which is what we really expect, and looks like where we started, (\ref{j}),
except here $\bf L$ is the orbital angular momentum, and this result is
not the same
as (\ref{jsplit}).
In fact, using the  result of Ref.\cite{allen}, (\ref{lf}) gives the $z$ component of the angular momentum as being proportional to $l$, the orbital topological number, as we expect. However, using the $L$ of (\ref{jsplit}), a plane polarized wave gives

\beq
L_z=\frac{E^2}{4\pi\om}({\bf k}\times{\bf r})_z\cos^2\Upsilon
\eeq
a result that is clearly wrong, since it should be zero.

In summary, the longstanding issue regarding the orbital and spin angular momentum of light has been addressed, and it was shown the culprit causing most of the problems was the surface term that was set to zero without adequate justification. Foregoing that approach, it was shown the antisymmetric part of the canonical energy momentum tensor leads naturally to the concept of spin. It was also shown symmetrizing the energy momentum tensor by discarding boundary terms is not always allowable and, in general, the antisymmetric energy momentum tensor must be retained. With this, the Lorentz covariant, gauge invariant spin vector was derived and shown to make sense.

\ed